\documentclass[aps,prl,10pt,twocolumn,showpacs,preprintnumbers,amsmath,amssymb,]{revtex4-1} 

\usepackage{float}
\usepackage{graphicx}
\usepackage{tikz}
\usepackage{wasysym} 
\usepackage{dcolumn}
\usepackage{color}
\usepackage{amsmath}
\usepackage[breaklinks,colorlinks,bookmarks=false,citecolor=blue,linkcolor=red,urlcolor=blue]{hyperref}

\renewcommand{\vec}[1]{{\mathbf #1}}

\begin{document}

\title{Diagnosing Fractionalization from the Spin Dynamics of $Z_2$  Spin Liquids\\ on the Kagome Lattice by Quantum Monte Carlo Simulations}

\author{Jonas Becker}
\affiliation{Institut f\"ur Theoretische Festk\"orperphysik, JARA-FIT and JARA-HPC, RWTH Aachen University, 52056 Aachen, Germany}
\author{Stefan Wessel}
\affiliation{Institut f\"ur Theoretische Festk\"orperphysik, JARA-FIT and JARA-HPC, RWTH Aachen University, 52056 Aachen, Germany}
\date{\today}

\begin{abstract}
Based on  large-scale quantum Monte Carlo simulations, we examine the dynamical spin structure factor of the Balents-Fisher-Girvin kagome lattice quantum spin-$1/2$ model, which is known to 
harbor an extended $Z_2$ quantum spin liquid phase. We use a correlation-matrix  sampling scheme
 combined with a stochastic analytic continuation method to resolve 
the spectral functions of this anisotropic quantum spin model with a three-site unit cell. Based on this approach, we monitor the spin dynamics throughout  
the  phase diagram  of this model,  from the $XY$-ferromagnetic region to the  $Z_2$ quantum spin liquid regime. In the latter phase, we identify a gapped two-spinon
continuum in the transverse scattering channel, which is faithfully modeled  by an effective spinon tight-binding model. Within the longitudinal  channel, we identify
 gapped vison excitations and exhibit indications for the translational symmetry fractionalization of the visons via an enhanced spectral periodicity. 
\end{abstract}

\maketitle
The search for quantum spin liquid (QSL)  states in frustrated magnets remains an active area of research in condensed matter physics~\cite{Balents10,Zhou17,Savary17}. 
As topologically ordered states of matter~\cite{Wen16}, gapped QSLs  exhibit long-ranged many-body entanglement and 
fractionalized excitations beyond one dimension. In order to identify QSL states in actual materials, spectroscopic measurements such as inelastic neutron scattering
constitute valuable diagnostic tools, e.g.,  by detecting scattering continua due to  deconfined fractionalized spin excitations or anyonic statistics~\cite{Morampudi17}.
Thus, it is important to obtain unbiased theoretical predictions for the
corresponding dynamical spin structure factor (DSSF) of fundamental microscopic models with gapped QSL  phases.  
A hallmark two-dimensional geometry in support of strong geometric frustration is the kagome lattice of corner-sharing triangles.
An unbiased theoretical characterization of kagome-lattice based QSL phases in terms of spectroscopic probes is  important in view of the compound
ZnCu${}_3$(OH)${}_6$Cl${}_2$ (herbertsmithite)~\cite{Fu15,Han16},  as well as Cu${}_3$Zn(OH)${}_6$FBr (Zn-doped barlowite), put forward recently as a possible spin-$1/2$ kagome gapped QSL candidate~\cite{Feng17,Wen17,Wei17}.
While indications of a gapped QSL ground state in the antiferromagnetic Heisenberg model on the kagome lattice have  been reported~\cite{Jiang08,Yan11,Nakano11,Depenbrock12},
other recent numerical results  indicate  that, instead, a  gapless QSL may be realized in this model~\cite{Iqbal13,Iqbal15,He17,Liao17}. 
This leaves the stability of a gapped topological QSL  open  for this  fundamental SU(2)-symmetric  model of kagome-lattice based quantum spin physics. 

\begin{figure}[t]
	\includegraphics{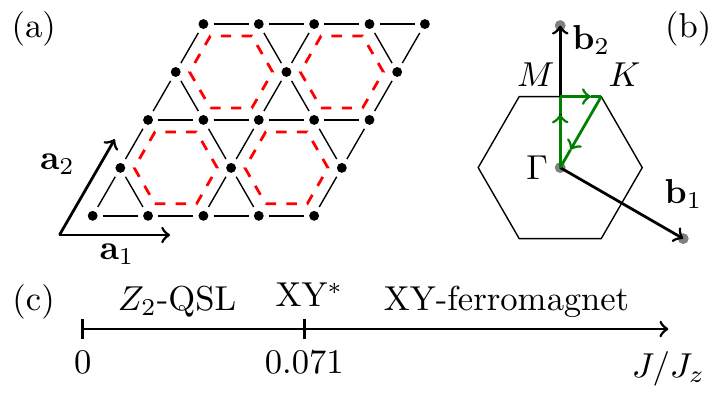} 
\caption{
(a) Kagome lattice with the nearest neighbor exchange (solid lines) and hexagonal cluster (dashed lines) terms of the BFG model. 
(b) First Brillouin zone (BZ) along with the path $\Gamma\rightarrow M\rightarrow K\rightarrow\Gamma$ (green lines).
(c) Ground state phase diagram of the BFG model.   
}\label{fig_sketch}	
\end{figure}

In this respect, it is comforting that  another, though anisotropic, spin-$1/2$ model by Balents, Fisher and Girvin (BFG) has  been established to harbor an extended gapped QSL on the kagome lattice~\cite{Balents02},  characterized as a $Z_2$ topologically ordered state~\cite{Wen91a, Wen91b}.
As such, this model hosts gapped deconfined spinons, i.e., spin-$1/2$ excitations,  in sharp contrast to conventional integer spin excitations such as magnons.  In terms of  anyonic statistics, the spinons in the BFG model are bosonic~\cite{Balents02}. 
In addition, the topological QSL phase of the BFG model  also exhibits gapped vortex excitations of an emerging odd-structured Ising gauge field~\cite{Wegner71,Fradkin78,Kogut79}, the so-called "visons"~\cite{Senthil00,Senthil01,Moessner01}. 
Within a resonating valence bond (RVB) description of the QSL ground state, 
a single vison excitation effects sign changes in 
the RVB superposition of singlet states across a semi-infinite line~\cite{Balents02}.
Carrying neither charge nor spin, visons do not couple directly to neutrons
but can be probed through their interaction with spinons~\cite{Punk14}.
Furthermore, 
their existence leads to a topology protected fourfold ground state degeneracy
in a system with periodic boundary conditions in both lattice directions~\cite{Balents02,Wen91a, Wen91b}. 
The BFG model, along with  several variants, was, indeed, shown to  exhibit 
a topological contribution to the ground state entanglement 
as well as symmetry-protected edge states due to nontrivial symmetry fractionalization~\cite{Sheng05, Isakov06,Isakov07, Isakov11, Isakov12,Wang17}. 
Part of this progress was possible since this model allows for sign-problem free, unbiased quantum Monte Carlo (QMC)  simulations on relatively large lattices. 
Hence, the BFG model is  well suited for probing the spin dynamics of $Z_2$ QSL states on the kagome lattice for gapped deconfined spinon and vison  excitations, based on unbiased QMC simulations.  

Thus, here we consider this basic model of an extended $Z_2$ QSL phase  on the kagome lattice to  study the corresponding DSSF using  advanced QMC methods. 
%
We  
identify  characteristic features of fractionalization
in the QSL phase and contrast them to the DSSF in the $XY$-ferromagnetically  ordered region of this model. 
The Hamiltonian  that we consider reads [cf. Fig.~\ref{fig_sketch}(a)]
$	H  = -J\sum\limits_{\langle j,j' \rangle} \left( S_j^+S_{j'}^- + S_j^-S_{j'}^+ \right)
		+\frac{J_z}{2} \sum\limits_{\hexagon} (S^z_{\hexagon})^2,$
in terms of a ferromagnetic ($J>0$)  nearest neighbor transverse spin exchange  and  longitudinal antiferromagnetic interactions of strength $J_z>0$ on all bonds within the hexagons  of the kagome lattice,  compactly expressed in terms of the
hexagonal  cluster terms, $S^z_{\hexagon}=\sum_{j\in \hexagon} S_j^z$.  In the following, we consider 
finite rhombic systems with $N_s=3L^2$ lattice sites and periodic boundary conditions along both  lattice directions $\mathbf{a}_1$, $\mathbf{a}_2$ in Fig.~\ref{fig_sketch}(a), 
with the unit cell distance fixed to  $a=1$. The first Brilliouin zone (BZ), in terms of the  reciprocal lattice vectors $\vec{b}_1$,  $\vec{b}_2$,  is shown in Fig.~\ref{fig_sketch}(b).
The Hamiltonian $H$ was shown in previous studies to harbor a QSL ground state for small values of $J$ that is separated from an $XY$-ferromagnetic region by a quantum critical point at $J/J_z=0.07076(1)$, which 
was identified as an $XY^*$ transition 
~\cite{Isakov11,Isakov12}, cf. Fig.~\ref{fig_sketch}(c). To formulate the DSSF of the three-sublattice kagome lattice, we 
denote by $\vec{S}_{i,\alpha}$  the spin at position $\vec{r}_{i,\alpha}$ on sublattice $\alpha$ ($=1,2,3$) in the $i$th unit cell ($i=1,...,L^2$) , and then obtain $3\times 3$ correlation matrices 
$S^{+-}_{\alpha,\beta}(\vec{k},\omega) = \int dt \, e^{-\mathrm{i}\omega t} \langle {S}^+_{\vec{k},\alpha}(t) {S}^-_{-\vec{k},\beta}(0) +  {S}^-_{\vec{k},\alpha}(t) {S}^+_{-\vec{k},\beta}(0)\rangle$
for the transverse,  and
$S^{zz}_{\alpha,\beta}(\vec{k},\omega) = \int dt \, e^{-\mathrm{i}\omega t} \langle {S}^z_{\vec{k},\alpha}(t) {S}^z_{-\vec{k},\beta}(0)  \rangle$
for the longitudinal 
channel respectively, where
$\vec{S}_{\vec{k},\alpha}=(1/L) \sum_i e^{-\mathrm{i} \vec{k}\cdot \vec{r_{i,\alpha} }} \vec{S}_{i,\alpha}$. In what follows, we examine, separately, the traces over the  correlation matrices in each channel, 
$S^{+-}(\vec{k},\omega) :=\sum_\alpha S^{+-}_{\alpha,\alpha}(\vec{k},\omega)$, and 
$S^{zz}(\vec{k},\omega) :=\sum_\alpha S^{zz}_{\alpha,\alpha}(\vec{k},\omega)$, respectively, as they contain the full summations over the correlation-matrix eigenvalues of the spectral functions at each fixed momentum transfer.

To calculate the spin spectral functions of the BFG model, we performed QMC simulations using the stochastic series expansion method~\cite{Sandvik99} for system size up to $L=18$.
The QMC sampling of this highly frustrated model in the strongly anisotropic QSL regime $J\ll J_z$ was improved by using a decoupling of $H$ in terms of four-site clusters~\cite{Louis04,Melko07,Wang17}, combined with directed loop updates~\cite{Syljuasen02, Alet05}.
In order to access ground state properties of the QSL,  the  temperature $T$ was tuned sufficiently below the vison excitation gap, as detailed below. 
To obtain the DSSF
from the QMC simulations, we  measured the corresponding transverse imaginary-time displaced spin-spin correlation functions and accessed the longitudinal correlations directly in Matsubara frequency space~\cite{Michel07,Michel07a},  using the stochastic analytic continuation method in the formulation of Ref.~\cite{Beach04} to obtain the  
spectral functions in real frequency space. 
We  performed the analytic continuations independently for the three correlation-matrix eigenvalues for each given momentum $\vec{k}$. These are obtained by  diagonalizing the $3\times 3$  correlation matrices in the imaginary-time domain, based on the binned QMC time-series data to estimate the corresponding covariances that enter the analytic continuation~\cite{Jarrell96,Beach04}.
The  spectral functions from the separately continued eigenvalues then yield $S^{+-}(\vec{k},\omega)$ and $S^{zz}(\vec{k},\omega)$ upon summation.
 
\begin{figure}[t]
\includegraphics[width=0.48\textwidth]{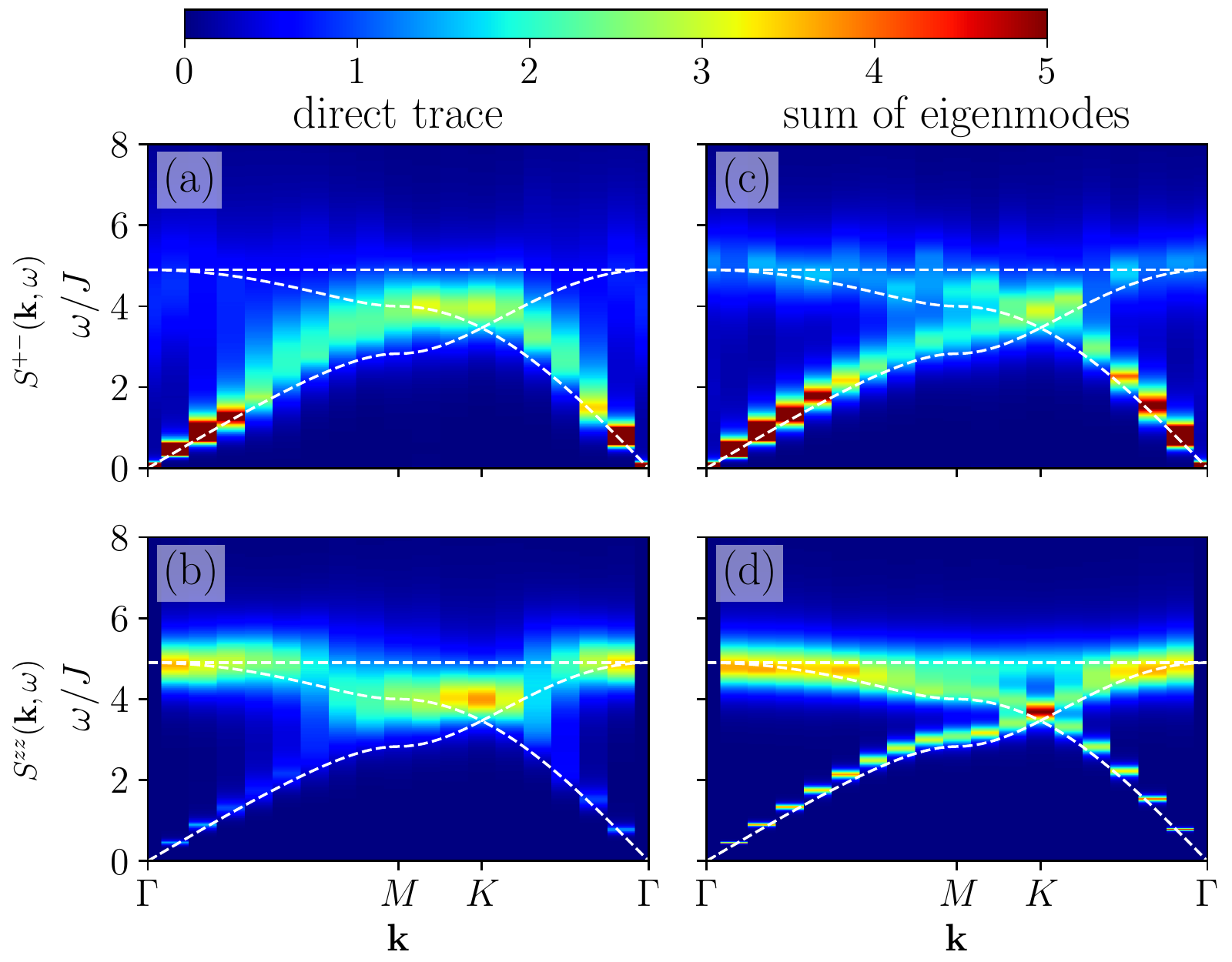}
\caption{
	DSSF $S^{+-}(\vec{k},\omega)$ (a), (c) and $S^{zz}(\vec{k},\omega)$ (b), (d) of the BFG model for $J_z=0$
along the BZ path $\Gamma\rightarrow M\rightarrow K\rightarrow\Gamma$, obtained at $T=0.03J$ on an $L=18$ lattice.
	(a) and (b)  show results obtained from analytic continuations of the direct correlation-matrix trace, and (c) and (d) show the summed  spectra from an eigenvalue decomposition.
Dashed lines are  magnon branches from LSWT. 
}\label{fig_xylimit}
\end{figure}

The benefits of this procedure can be demonstrated by first considering the $XY$-ferromagnetic limit $J_z=0$ of the BFG model. In this limit, the system has an $XY$-ferromagnetic ground state (a superfluid phase in the bosonic formulation of the BFG model), which spontaneously breaks its residual U(1) symmetry. Figure~\ref{fig_xylimit}  shows  $S^{+-}(\vec{k},\omega)$  and $S^{zz}(\vec{k},\omega)$ along a high-symmetry BZ path [cf.~Fig~\ref{fig_sketch}(b)], and
compares the spectra from the analytic continuation  of the direct trace (a), (b) to the summation over the separately continued correlation matrix eigenvalues (c), (d).  
Also shown in Fig.~\ref{fig_xylimit} are the three magnon branches from  linear spin-wave theory (LSWT) for the $XY$-ferromagnet on the kagome lattice,  obtained 
in the Holstein-Primakoff representation from a bosonic Bogoliubov matrix diagonalization~\cite{Avery74} in a rotated reference frame~\cite{Stephanovich11}. 
This comparison shows that the eigenvalue decomposition allows for a higher resolution of the distinct modes than the analytic continuation of the direct trace, 
since the single continuation of the direct trace fails to discern  close-by spectral features in frequency space that are due to distinct eigenvalues.  
In particular,  the transverse channel is dominated by the low-energy magnon branch, the Goldstone soft-mode that results from the spontaneous U(1) symmetry breaking. 
The broad continuum above the lowest magnon branch indicates multimagnon excitations.
The Goldstone mode also contributes to $S^{zz}(\vec{k},\omega)$  but with a much lower spectral weight that vanishes towards the $\Gamma$ point. 
A larger spectral  weight is supported by  the magnon branches at elevated energies of  $\omega\approx 4J$, 
which the eigenvalue decomposition allows us to separate [cf., e.g.,  the $K$ point in Fig.~\ref{fig_xylimit}(d)]. The exactly flat optical magnon mode from LSWT is  found to be, at most, weakly dispersive in QMC calculations. 

\begin{figure}[t]
\includegraphics[width=0.5\textwidth]{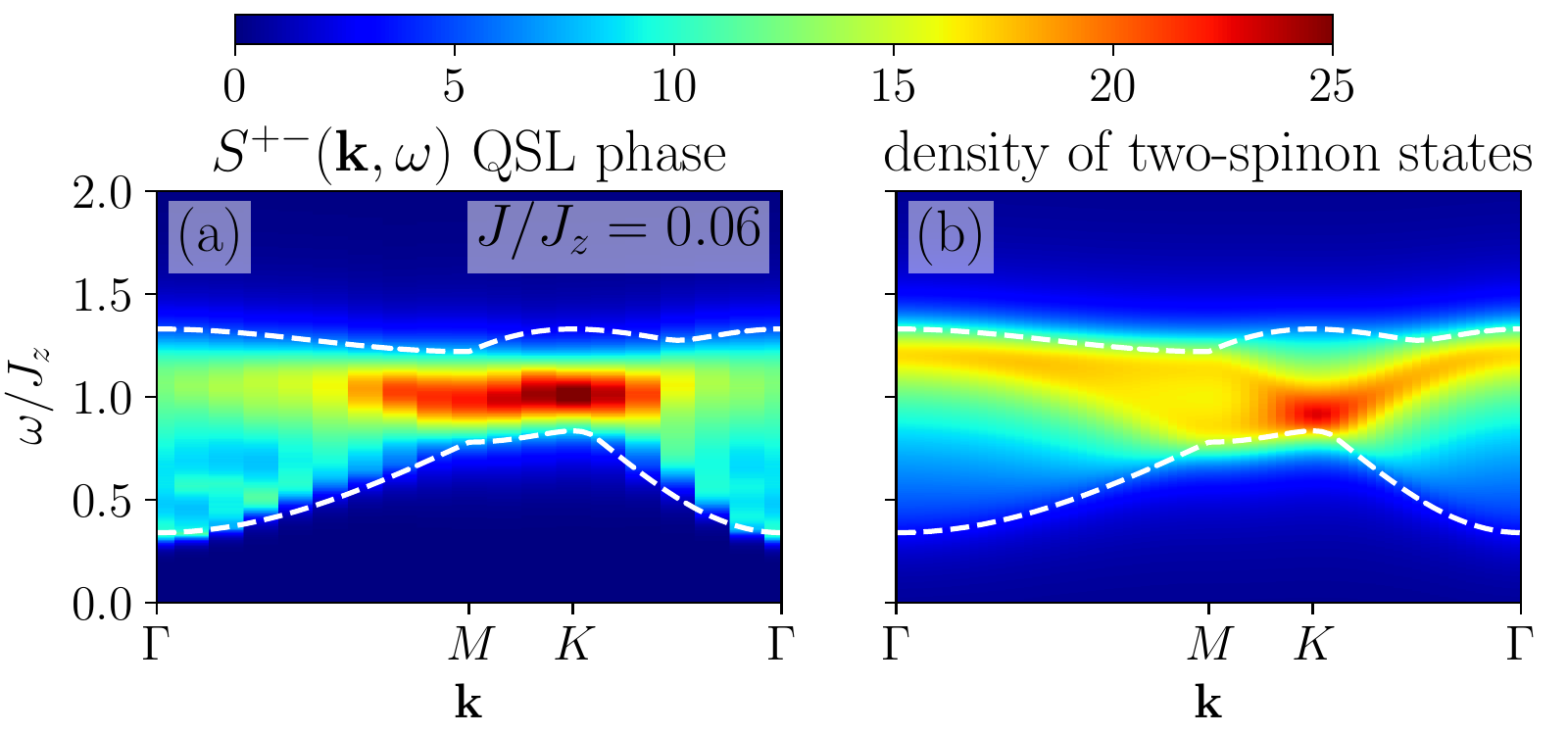}
\caption{
(a) DSSF $S^{+-}(\vec{k},\omega)$ of the BFG model for   $J/J_z=0.06$ within the QSL regime, 
along the BZ path $\Gamma\rightarrow M\rightarrow K\rightarrow\Gamma$, obtained at $T=0.002J_z$ on an $L=18$ lattice. 
(b) Density of two-spinon states within the tight-binding model with $t=0.055J_z$, and a Lorentzian $\delta$-function broadening of half-width $2t$.
Dashed lines in both panels show the lower and upper threshold of the two-spinon continuum within the tight-binding model. 
}\label{fig_QSL}
\end{figure}

The well-defined low-energy magnon excitations in the $XY$-ferromagnetic regime contrast strongly to the spin dynamics observed within the QSL regime, which we examine next.
The QMC result for $S^{+-}(\vec{k},\omega)$ at a value of $J/J_z=0.06$ is shown in  Fig.~\ref{fig_QSL}(a) (at even smaller  $J/J_z$, the QMC updates are much less efficient). We observe a broad continuum of gapped excitations, 
with a minimum gap at the $\Gamma$ point of about $0.4J_z$. In the strong $J_z$ regime, the low-energy configurations are characterized by states with $S^z_{\hexagon}=0$
on all hexagons, and a local spin flip creates two neighboring hexagons with $S^z_{\hexagon}=1$. Because of the transverse exchange $J>0$, these  defects delocalize  over the
triangular lattice of hexagons, forming two spatially separated spin-$1/2$ spinon excitations~\cite{Balents02}. To quantify the  two-spinon contribution to $S^{+-}(\vec{k},\omega)$,   
we  use  a tight-binding model that treats the spinons as free particles, with a dispersion relation $\epsilon_{\vec{k}}=J_z/2+\epsilon_t(\vec{k})$, where $J_z/2$ 
is the local energy cost of a single spinon, and
$\epsilon_t(\vec{k})=-2t[\cos(\vec{a_1}\vec{k})+\cos(\vec{a_2}\vec{k})+\cos(\vec{a_2}\vec{k}-\vec{a_1}\vec{k})]$
the triangular lattice tight-binding dispersion with $t\sim J$ the nearest-neighbor hopping amplitude, fitted to the bandwidth of the QMC continuum. 
In this approximation, $S^{+-}(\vec{k},\omega)\approx \frac{4\pi}{L^2} \sum_{\vec{k}'} \delta(\omega-\epsilon_{\vec{k}'}-\epsilon_{\vec{k}-\vec{k}' })$
is given by the density of two-spinon states, and interaction effects may be  accounted for phenomenologically by a Lorentzian $\delta$-function broadening~\cite{Kourtis16,Huang17}. 
This approximation is shown in Fig.~\ref{fig_QSL}(b).
Comparisons of the spectra at various fixed momenta are also available~\cite{sm}.
The tight-binding model captures the spectral support of the continuum and  the enhanced spectral weight near the BZ corners ($K$ points) at $\omega\approx J_z$. 
The  continuum  in $S^{+-}(\vec{k},\omega)$  provides a clear signature for deconfined spinons in the QSL phase, in strong contrast to the sharp gapped magnon excitations observed, e.g, in quantum disordered spin-dimer systems  or  valence bond ordered states~\cite{Lohoefer15,Lohoefer17,Ma18}.  

\begin{figure}[t]
\includegraphics[width=0.49\textwidth]{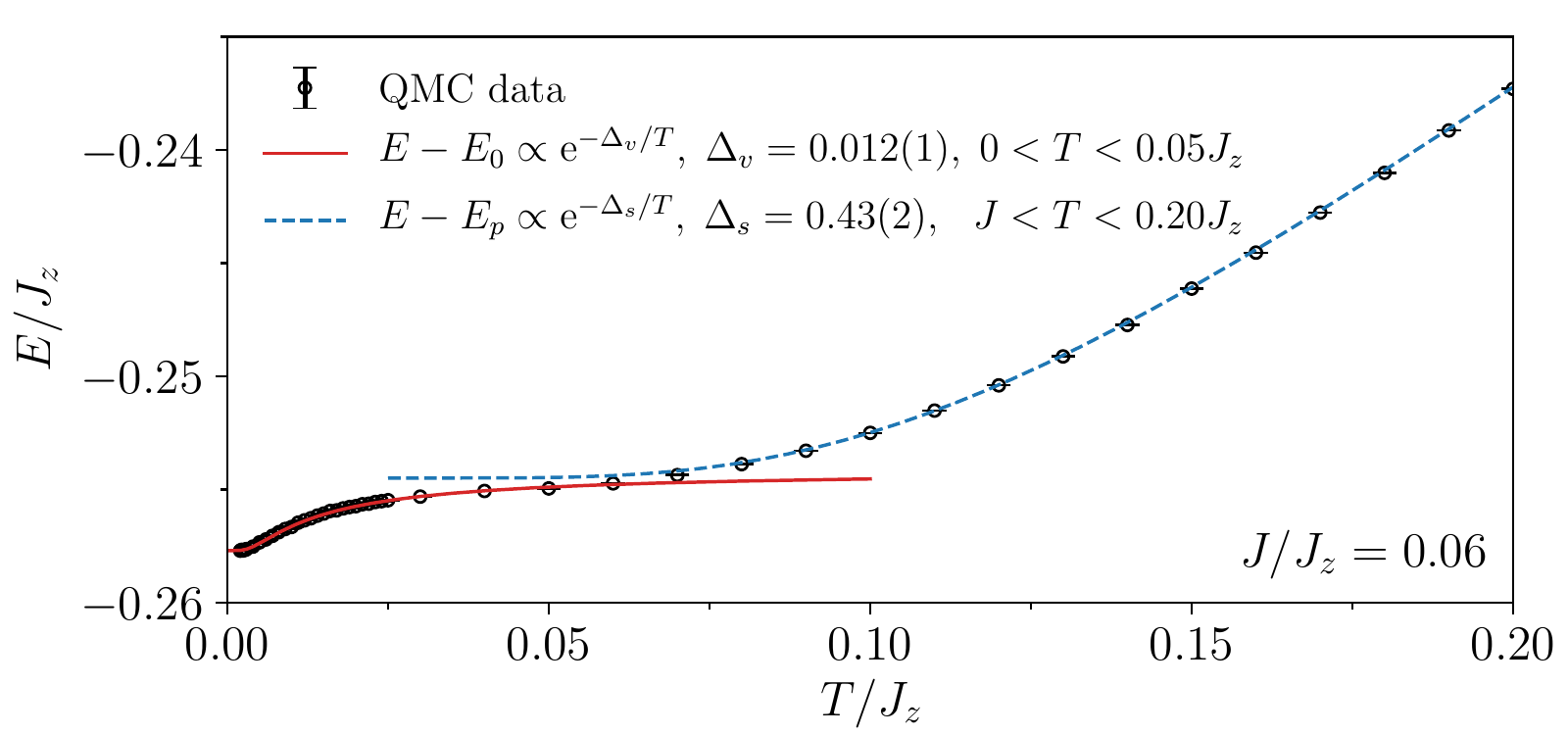}
\caption{
$T$ dependence of the energy $E$ of the BFG model for $J/J_z=0.06$, along with exponential fits to  activated behavior atop the ground state energy $E_0$ and the energy
plateau $E_p\approx -0.255 J_z$ at $T\approx J$ on an $L=18$ lattice.
}\label{fig_E}
\end{figure}

\begin{figure}[t]
\includegraphics[width=0.5\textwidth]{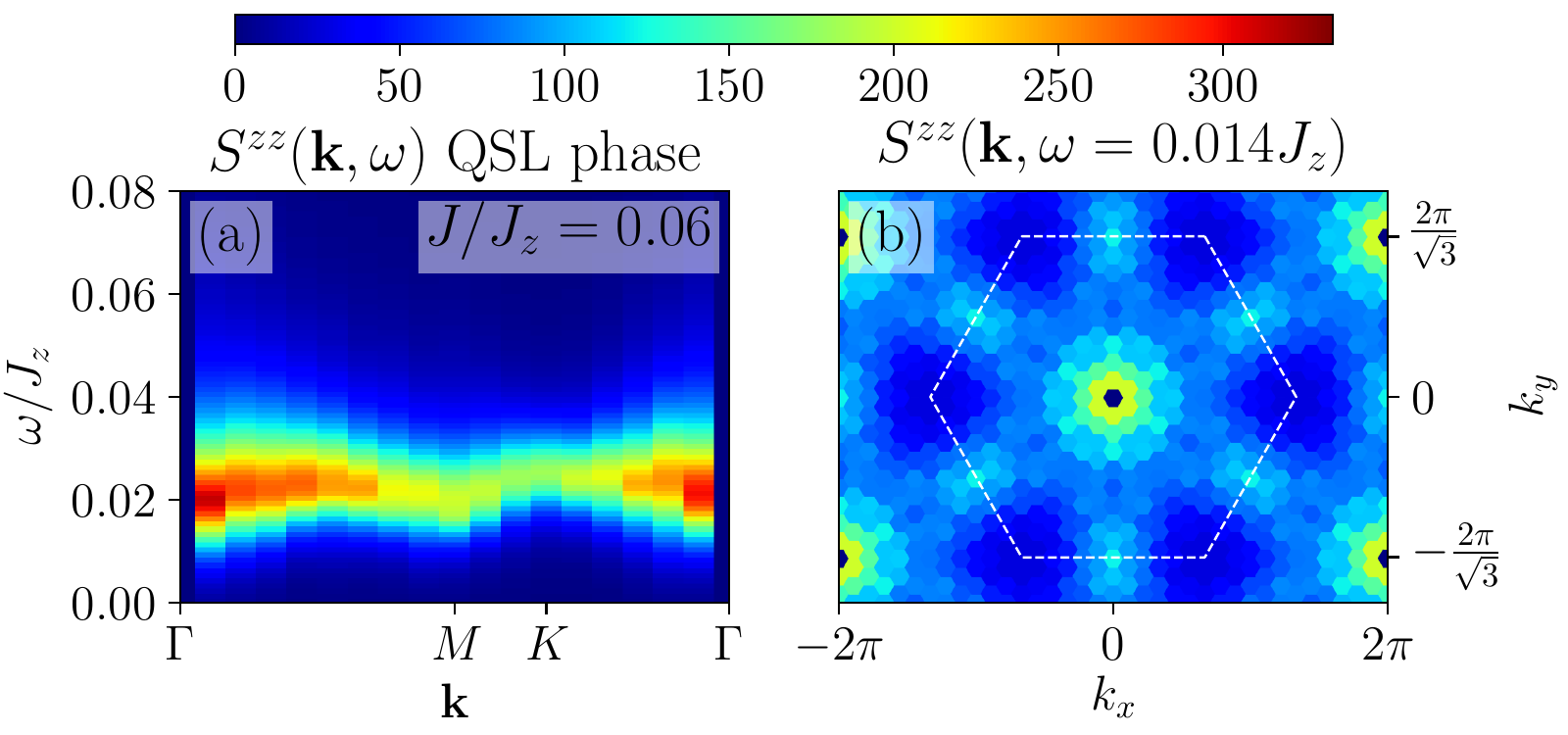}
\caption{
DSSF $S^{zz}(\vec{k},\omega)$ of the BFG model for  $J/J_z=0.06$ within the QSL regime  at $T=0.002J_z$ on an $L=18$ lattice,
(a) along the BZ path $\Gamma\rightarrow M\rightarrow K\rightarrow\Gamma$, 
and (b) at $\omega=0.014J_z$. The  hexagon in (b) denotes the BZ. 
}\label{fig_Szz}
\end{figure}

The  two-spinon gap from Fig.~\ref{fig_QSL} corresponds to  the  activated $T$ dependence of the energy $E$ atop the spin paramagnetic plateau~\cite{Isakov07}
at $T\approx J$, cf. the exponential fit to the data in Fig.~\ref{fig_E}, which yields $\Delta_s\approx 0.43(2) J_z$. The  activated behavior of $E$ at even lower temperatures, $T\ll J$, seen in  Fig.~\ref{fig_E}, instead, arises  from  the thermal proliferation of vison-pair excitations~\cite{Isakov07},
and from fitting $E(T)$ to an exponential form,  we estimate a corresponding two-vison gap of $\Delta_v\approx 0.012(1)J_z$. 
In the quantum-dimer model limit of the BFG model, for $J/J_z\rightarrow 0$, the equal-time vison-vison correlations are given by a string operator of $S^z_j$ operators~\cite{Balents02}, and correspondingly, two-vison excitations may be probed through the longitudinal channel $S^{zz}(\vec{k},\omega)$. 
The QMC result for $S^{zz}(\vec{k},\omega)$ for $J/J_z=0.06$ is shown in Fig.~\ref{fig_Szz}. From the BZ path data  in the left panel, we identify an excitation gap at the $\Gamma$ point 
 in accord with the above estimate. Thus, to obtain the spectral functions, we performed the QMC simulations at a temperature of $T=0.002J_z$, below the estimated two-vison gap. 
Based on our data, we cannot discern vison bound states below the continuum. Indications for such  bound states were reported for a related quantum dimer model on the triangular lattice~\cite{Laeuchli08}, for which, also, the  single-vison dispersion is available~\cite{Ivanov04},    based on which the spectral support of the two-vison continuum could, thus, be constructed and separated  from  the   bound state.

Upon closer inspection, one identifies, in Fig.~\ref{fig_Szz}(a), a  low-energy structure (at the two-vison gap energy)  in  $S^{zz}(\vec{k},\omega)$  at the $M$ point. This  is seen more explicitly in Fig.~\ref{fig_Szz}(b), which shows  $S^{zz}(\vec{k},\omega)$ at a constant energy $\omega=0.014J_z$ (similar cuts at other fixed energies are also available~\cite{sm}). This  repeating structure  at the $M$ points of the BZ (at momenta $\pm\vec{b}_1/2$,  $\pm\vec{b}_2/2$, and $\pm(\vec{b}_1+\vec{b}_2)/2$) is in accord with  an enhanced spectral periodicity
that is expected in the continuum of the two-vison states, based on the crystal momentum fractionalization of the vison excitations in  a $Z_2$ QSL on the kagome lattice, i.e., anticommuting translation operators $T_1$ and $T_2$,  $T_1 \:T_2=- T_2 \:T_1$, acting on  the single vison excitations along the two lattice directions ~\cite{Wen02,Huh11,Essin13,Qi15a,Qi15b,Qi16,Cheng16}.
For  gapped  $Z_2$ QSL states, such a spectral periodicity  was shown to  provide  a spectroscopic  diagnostic of  crystal momentum fractionalization of anyon excitations~\cite{Wen02, Essin14}. The low-energy spectral weight observed in  $S^{zz}(\vec{k},\omega)$  at the $M$ points suggests that the QSL state in the considered parameter regime 
of the BFG model can be driven towards a valence bond solid  instability upon adding an appropriate perturbation that couples to valence bond fluctuations at the $M$ points
%
~\cite{Xu09, Huh11}. 
An explicit investigation of such a scenario is provided in Ref.~\cite{Sun18}. 

\begin{figure}[t]
\includegraphics[width=0.5\textwidth]{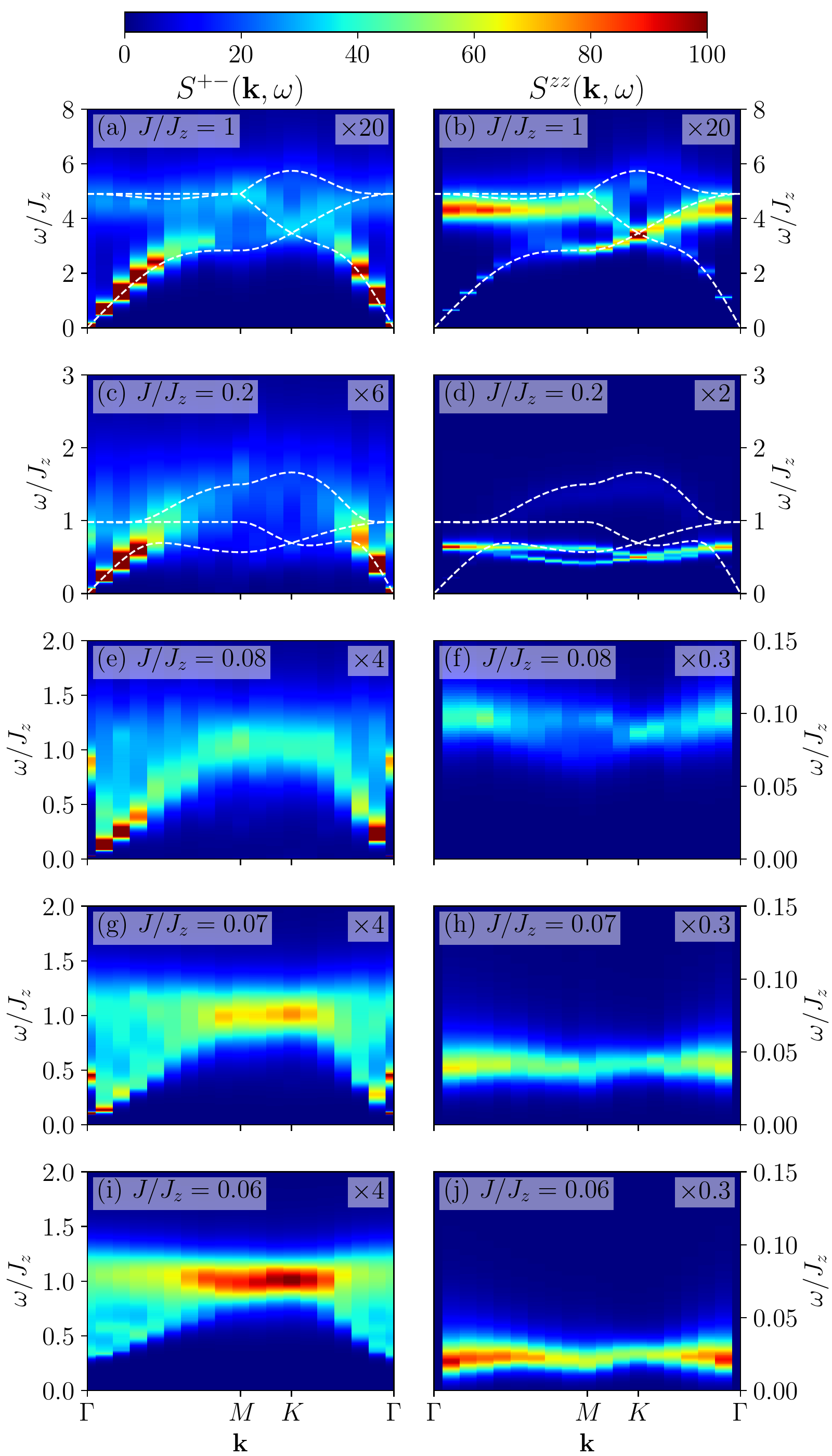}
\caption{
DSSF $S^{+-}(\vec{k},\omega)$ (left panels) and $S^{zz}(\vec{k},\omega)$ (right panels) along the BZ path
$\Gamma\rightarrow M\rightarrow K\rightarrow\Gamma$  for different ratios $J/J_z$ from QMC simulations at temperatures $T$
near ${J}/({2L}) $ on an $L=18$ lattice. 
To fit to a common scale, the intensities were multiplied by individual factors, which are provided in the upper right corner of each panel separately. Dashed lines are  magnon branches from LSWT. 
}\label{fig_all}
\end{figure}

Finally, we monitor, in Fig.~\ref{fig_all}, the evolution of the DSSF upon approaching the QSL phase from the $XY$-ferromagnetic regime. In the transverse channel, we observe  (i) a progressive broadening of the lowest magnon branch at high energies upon decreasing $J/J_z$, along with (ii) a reduced spin-wave velocity (as estimated by the slope of  the magnon branch  near the $\Gamma$ point), and  (iii) a redistribution of spectral weight to preform the  characteristic shape of the spinor-continuum in the QSL phase. In the longitudinal channel, 
the characteristic energy scale of the predominant optical modes drops rapidly with decreasing  $J/J_z$, an effect that is also qualitatively captured by LSWT. We also observe a strong suppression in 
 the spectral weight of the  lowest magnon mode. Upon entering the  QSL phase, the other two branches merge, and a gap minimum is  formed at  the $\Gamma$ point. 

Our results demonstrate the feasibility of probing  fractionalization in QSL states from the spin dynamics  
of microscopic models on relatively large lattices 
using unbiased numerical methods, such as QMC simulations.
It will be intriguing to compare  our results to the spin dynamics of other kagome-lattice based QSL states, e.g., with an even Ising gauge structure, which are also accessible to QMC approaches~\cite{Wang17a}.
Furthermore, alternative numerical approaches such as those used in Refs.~\cite{Gohlke17, Winter17} can also 
probe the spin dynamics of QSL states beyond the realms of QMC methods.

\noindent

We thank  Z. Y. Meng and Y. Qi for sharing their valuable insight on crystal symmetry fractionalization, and for communicating results from a related study~\cite{Sun18}.
We also thank  F. Hassler, A.M. L\"auchli, and F. Pollmann
for useful discussions.
Furthermore, we acknowledge support by the Deutsche Forschungsgemeinschaft (DFG) under Grants No. FOR 1807 and No. RTG 1995, and thank the IT Center at RWTH Aachen University and the JSC J\"ulich for access to computing time through JARA-HPC.

\newpage

\clearpage
\section{Supplemental Material}
\setcounter{figure}{0}  
\renewcommand{\thefigure}{S\arabic{figure}}

\begin{figure}[H] \centering
\includegraphics[width=\columnwidth]{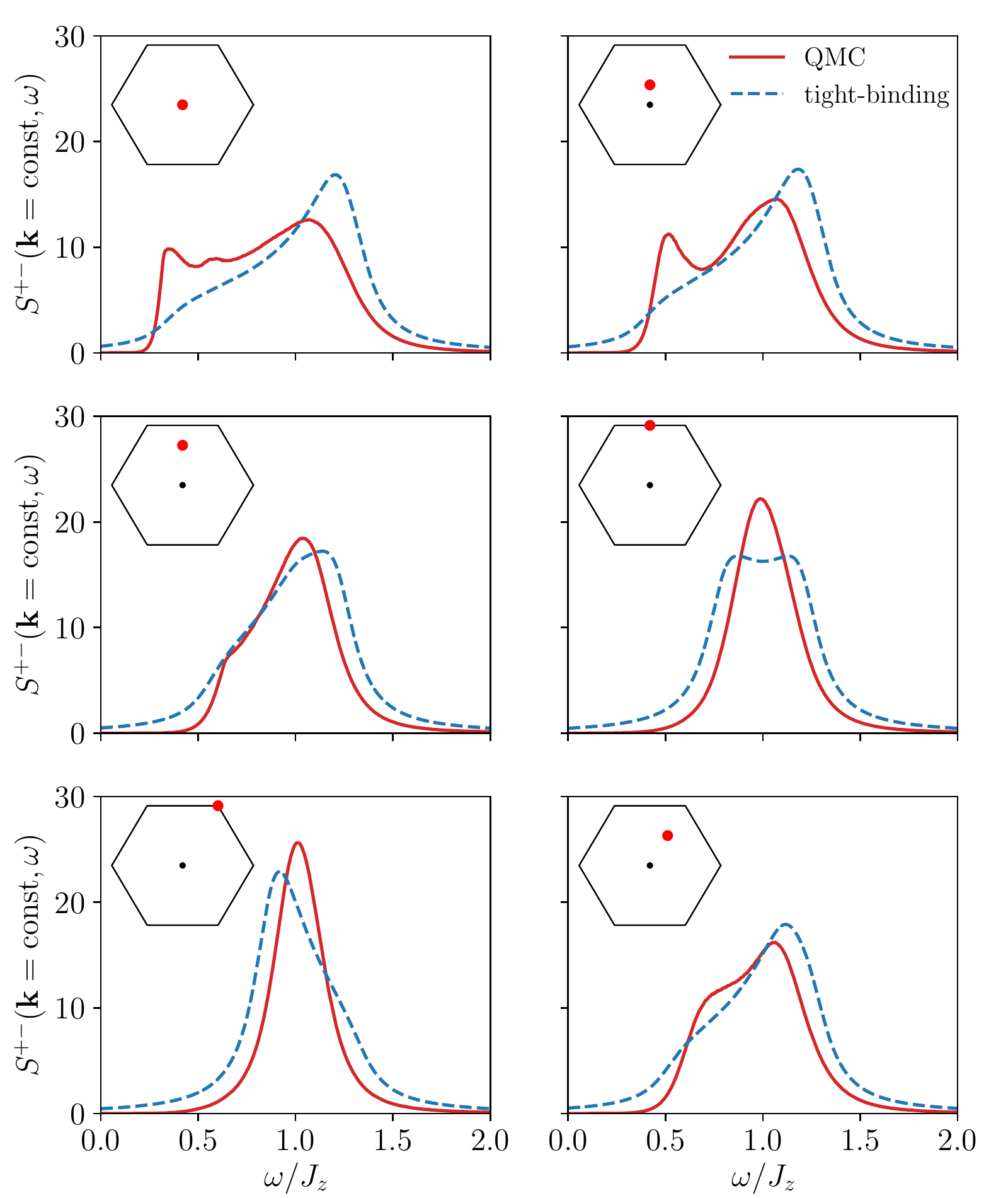}
\caption{
Dynamical spin structure factor  in the transverse channel $S^{+-}(\vec{k},\omega)$ at different fixed momenta $\vec{k}$, indicated by the red dots in the BZ insets,
as obtained from QMC simulations of the BFG model in the QSL phase at $J/J_z=0.06$, for
$T=0.002J_z$ on the $L=18$ lattice (solid lines). Also shown by dashed lines are the corresponding
  two-spinon spectra from  the tight-binding model  with $t=0.055J_z$, and a Lorentzian $\delta$-function broadening of half-width $2t$.}
\label{fig_kcuts}
\end{figure}

\begin{figure}[H] \centering
\includegraphics[width=\columnwidth]{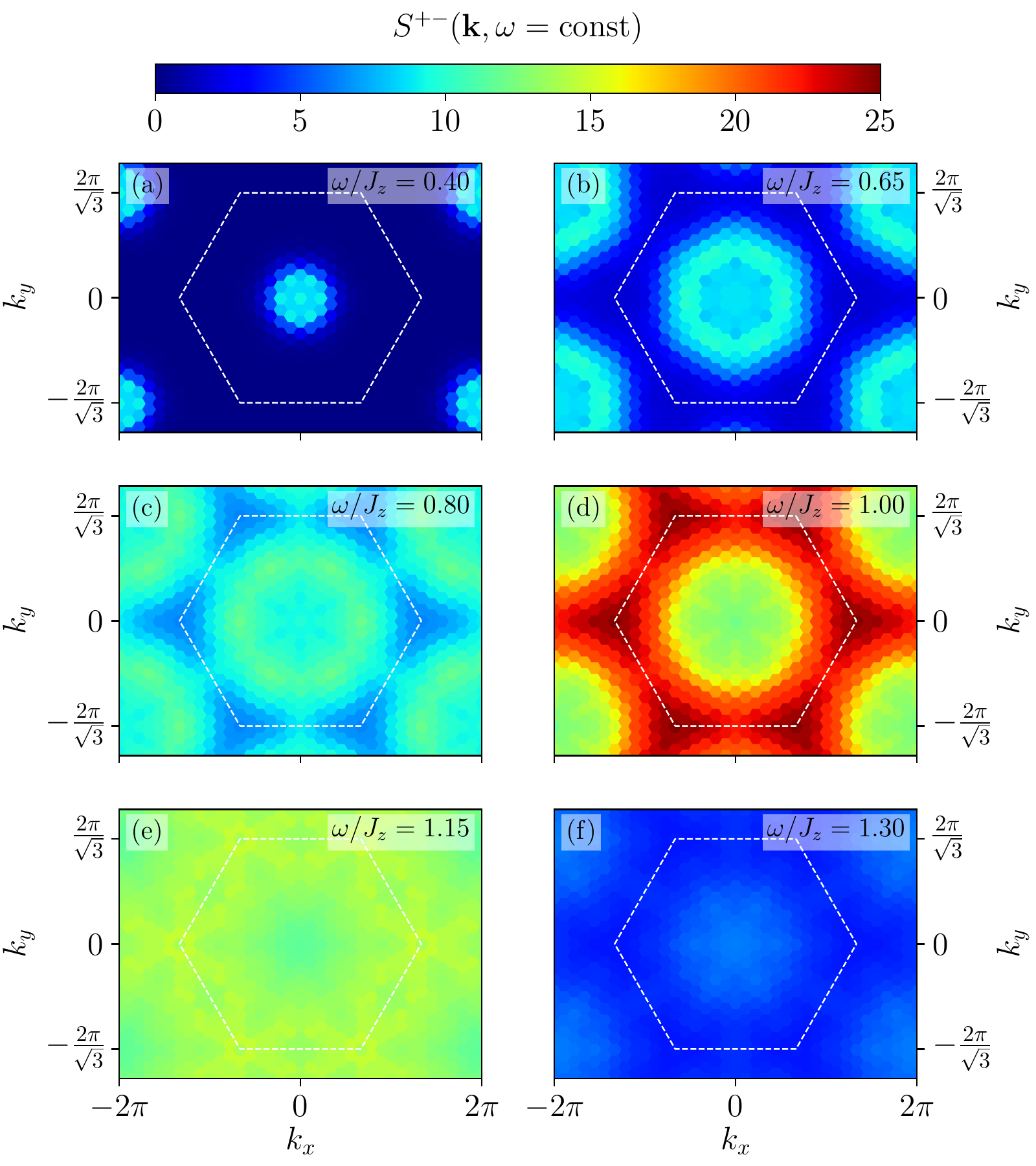}
\caption{
Dynamical spin structure factor in the transverse channel $S^{+-}(\vec{k},\omega)$ at different fixed energies $\omega$,
as obtained from QMC simulations of the BFG model in the QSL phase at $J/J_z=0.06$, for
$T=0.002J_z$ on the $L=18$ lattice.
}
\label{fig_energycuts_xy}
\end{figure}

\begin{figure}[H] \centering
\includegraphics[width=\columnwidth]{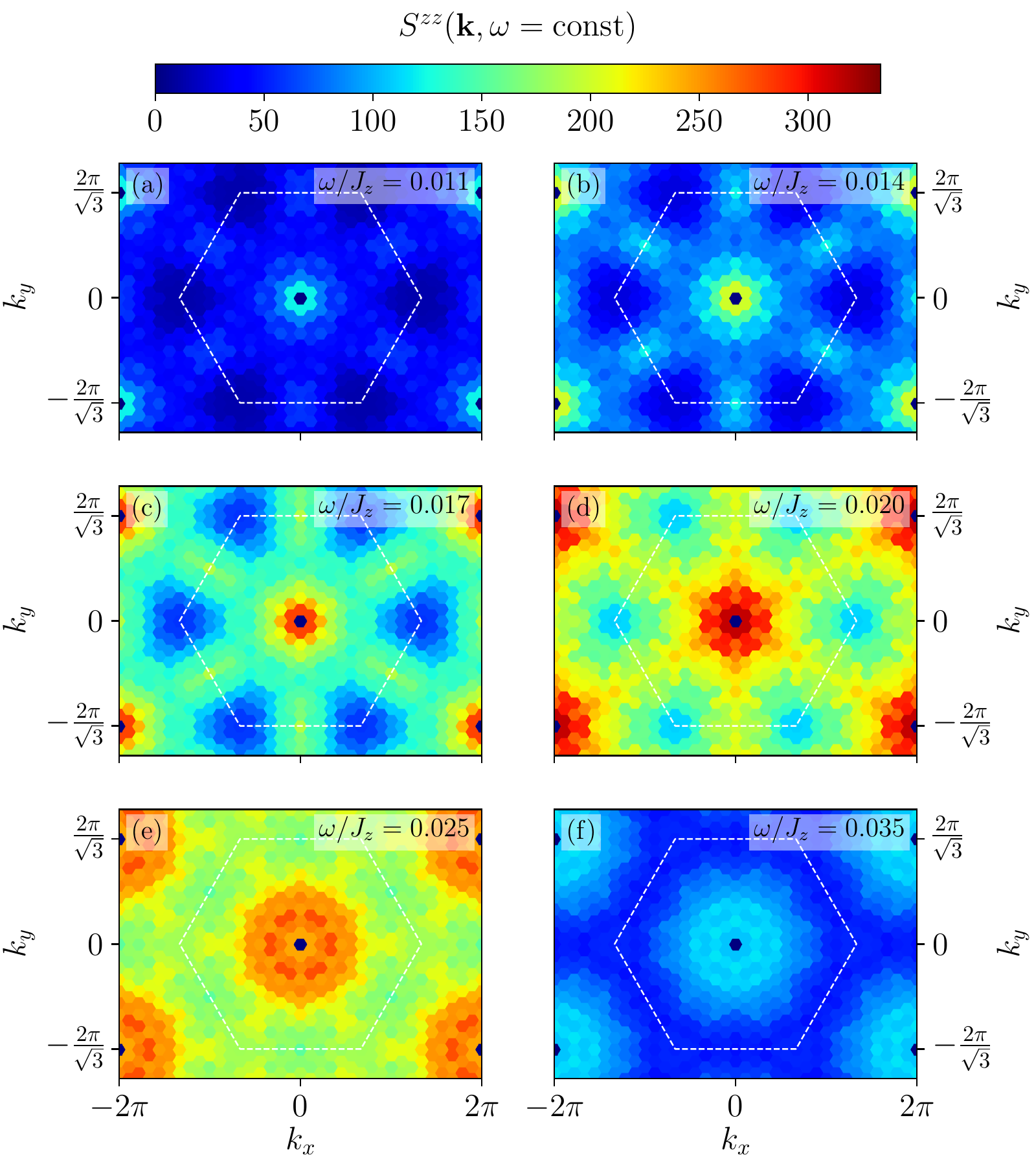}
\caption{
Dynamical spin structure factor in the longitudinal channel $S^{zz}(\vec{k},\omega)$ at different fixed energies $\omega$,
as obtained from QMC simulations of the BFG model in the QSL phase at $J/J_z=0.06$, for
$T=0.002J_z$ on the $L=18$ lattice.
}
\label{fig_energycuts_zz}
\end{figure}

Figure~\ref{fig_kcuts} shows the  dynamical spin structure factor of the BFG model in the transverse channel, $S^{+-}(\vec{k},\omega)$, at different fixed momenta $\vec{k}$, for $J/J_z=0.06$ within the QSL phase, along with a comparison to the two-spinon spectra from  the tight-binding model introduced in the main text.
Furthermore, Figs.~\ref{fig_energycuts_xy} and \ref{fig_energycuts_zz} show the dynamical spin structure factor in the transverse and longitudinal channels, $S^{+-}(\vec{k},\omega)$, and  $S^{zz}(\vec{k},\omega)$, respectively, at different fixed energies $\omega$ for the same value of $J/J_z$ from the QSL regime.

\end{document}